# Disentangling Strain and $Ti^{3+}$ Contributions to the Anomalous Hall Effect in Epitaxial $RuO_2$ Films

Seung Gyo Jeong[1,2,*], Anand Santhosh[1], Seungjun Lee[3,4], Sharup Sheikh[5], Uditha M. Jayathilake[5], Tien-Lin Lee[6], Tony Low[3], Alexander X. Gray[5], Scott A. Chambers[7], and Bharat Jalan[1,*]

*[1]Department of Chemical Engineering and Materials Science, University of Minnesota−Twin Cities, Minneapolis, Minnesota 55455, USA*

*[2]Department of Physics, Hankuk University of Foreign Studies, Yongin 17035, Republic of Korea*

*[3]Department of Electrical and Computer Engineering, University of Minnesota−Twin Cities, Minneapolis, Minnesota 55455, USA*

*[4]Department of Applied Physics, Kyung Hee University, Yongin 17104, Republic of Korea*

*[5]Department of Physics, Temple University, Philadelphia, PA, 19122 USA*

*[6]Diamond Light Source Ltd., Oxfordshire OX11 0DE, United Kingdom*

*[7]Physical and Computational Sciences Directorate, Pacific Northwest National Laboratory, Richland, WA 99354, USA*

[*]Corresponding authors: seunggyo625@hufs.ac.kr; bjalan@umn.edu

**Abstract**

The anomalous Hall effect (AHE) reported in epitaxial $RuO_2/TiO_2$ has been attributed to a strain-stabilized magnetic state, but strain can be entangled with Ru-Ti intermixing and interfacial charge redistribution, which can produce $Ti^{3+}$ and potentially localized moments. Here, we disentangle these contributions using Ti-alloyed $RuO_2$ heterostructures in which abundant $Ti^{3+}$ states are retained while epitaxial strain is independently changed. The strained and relaxed films have nearly identical Ti/Ru compositions and comparable $Ti^{3+}$ fractions, yet a pronounced nonlinear AHE appears only in the coherently strained film, while the relaxed heterostructure exhibits an almost linear Hall response. Spectroscopic ellipsometry further shows that the AHE-active strained state is accompanied by a reconstruction of the itinerant electronic response, including enhanced metallicity and longer carrier relaxation times. These results show that $Ti^{3+}$ formation alone is insufficient to generate the anomalous Hall state and identify epitaxial strain, through its modification of the itinerant $RuO_2$ electronic structure, as the dominant control parameter.

## Introduction

Epitaxial interfaces provide a powerful route to stabilize electronic and magnetic states absent in bulk materials [1-5], but the same interface can simultaneously impose lattice strain, drive charge transfer, and promote atomic reconstruction. Separating these coupled effects is therefore essential for identifying the microscopic origin of an interfacial phase. This challenge is particularly acute for $RuO_2$, which has emerged as a prominent candidate for altermagnetism [6,7]. Epitaxial $RuO_2$ exhibits a broad range of magnetic responses [8-42], despite bulk $RuO_2$ being largely regarded as nonmagnetic [43-49]. In particular, fully strained ultrathin ($\leq$ 4 nm) $RuO_2/TiO_2(110)$ grown by hybrid molecular beam epitaxy (MBE) remains metallic to atomic-scale thickness and exhibits symmetry-breaking or magnetic signatures in optical second-harmonic generation [19], magneto-optical measurements [19], spin-resolved angle-resolved photoemission spectroscopy (ARPES) [39], polarized neutron reflectometry (PNR) [50], and the anomalous Hall effect (AHE) [20]. Epitaxial strain has been proposed as the driving mechanism [19,20,38,39], but the $RuO_2/TiO_2$ interface also provides an alternative microscopic route: charge redistribution, and potentially Ru-Ti intermixing, can create $Ti^{3+}$ states that introduce additional d electrons and may generate localized magnetic moments [51-54]. A central unresolved question is therefore whether the AHE reflects a strain-reconstructed itinerant state of $RuO_2$ or instead originates from Ti-related interfacial electronic reconstruction.

In this Letter, we design an experiment that separates these two possibilities. We first use angle-dependent hard X-ray photoemission spectroscopy to establish the presence of $Ti^{3+}$ at the $RuO_2/TiO_2$ interface. We then intentionally alloy Ti into $RuO_2$ ($Ti_xRu_{1-x}O_2$), creating a high density of $Ti^{3+}$ states throughout the ultrathin conducting layer, while independently tuning its epitaxial boundary condition by inserting a strain-relaxed $SnO_2$ buffer. The resulting strained and relaxed $Ti_xRu_{1-x}O_2$ films retain nearly identical compositions and $Ti^{3+}$ populations, providing a direct test of whether $Ti^{3+}$ is sufficient to produce the AHE. Despite comparable $Ti^{3+}$ populations, the AHE remains robust in the coherently strained film but is strongly suppressed with strain relaxation. Spectroscopic ellipsometry further shows that strain simultaneously reconstructs the itinerant carrier response. These experiments collectively establish epitaxial strain, rather than the mere presence of $Ti^{3+}$ states, as the dominant control of the anomalous Hall state in ultrathin $RuO_2/TiO_2$-based heterostructures.

**Results and Discussion**

We first investigated the presence of $Ti^{3+}$ at $RuO_2/TiO_2$ interface. Angle-dependent hard X-ray photoemission spectroscopy (HAXPES) on a 1.4 nm $RuO_2/TiO_2(110)$ sample (Fig. S1 [55]) provides depth-sensitive access to the buried interface. Although conventional laboratory XPS does not resolve a $Ti^{3+}$ component because of its more limited probing depth [56], HAXPES reveals a small but clear $Ti^{3+}$ spectral weight at grazing take-off angles. Quantitative analysis is consistent with a ~1.4 nm-thick interfacial region containing ~16 % $Ti^{3+}$. Importantly, modeling of the $Ti^{3+}/Ti^{4+}$ ratio gives comparably good agreement for two distinct microscopic scenarios: $Ti^{3+}$ confined to an interfacial $TiO_2$ layer or $Ti^{3+}$ incorporated in an interfacial $Ti_xRu_{1-x}O_2$ layer. The spectroscopy therefore establishes $Ti^{3+}$ formation but does not uniquely determine whether it arises from charge transfer or local Ru-Ti intermixing. In bulk, the electron affinity of insulating $TiO_2$ (~4 eV) [57] and the work function of metallic $RuO_2$ (~5 eV) [58,59] would suggest a sizable interfacial barrier, whereas ultrathin-interface electrostatics, including strain-induced dipoles and band bending [58], can substantially modify the local potential landscape. Prior multislice electron ptychography of hybrid-MBE-grown $RuO_2/TiO_2$ did not resolve appreciable oxygen vacancies or Ru-Ti intermixing within experimental resolution [20,57]. The microscopic location of $Ti^{3+}$ therefore remains open, making it essential to determine experimentally whether $Ti^{3+}$ itself is sufficient to account for the AHE.

We therefore constructed a controlled test of the $Ti^{3+}$ hypothesis by growing two epitaxial

$Ti_xRu_{1-x}O_2$ heterostructures with contrasting strain states by hybrid MBE (see Supplemental Material [55]) [19,20,38,39,56]. Ti alloying intentionally creates abundant Ti-derived reduced states throughout the ultrathin $RuO_2$-based conducting layer; it is used here not as an exact microscopic replica of the buried interface, but as a stringent test of whether a large $Ti^{3+}$ concentration is sufficient to generate the AHE. Strain was changed independently by inserting a $SnO_2$ buffer, yielding $Ti_xRu_{1-x}O_2$/2 nm $TiO_2$/$TiO_2$(110) and $Ti_xRu_{1-x}O_2$/2 nm $TiO_2$/12 nm $SnO_2$/$TiO_2$(110). The corresponding bulk lattice mismatches are summarized in Fig. S2 [55,60,61]. The same 2 nm $TiO_2$ layer was included directly beneath $Ti_xRu_{1-x}O_2$ in both structures to keep the immediate interface chemistry as similar as possible. Without $SnO_2$, $Ti_xRu_{1-x}O_2$ is coherently constrained by $TiO_2$(110); with $SnO_2$, the relaxed buffer imposes a substantially different epitaxial boundary condition. For $RuO_2$, density functional theory (DFT) predicts that the strain associated with $SnO_2$ (2.6% tensile along [001] and 5.5% tensile along $[1\bar{1}0]$)) is nonmagnetic, whereas full coherency to $TiO_2$ (−4.7% compressive along [001] and 2.3% tensile along [10]) stabilizes a magnetic ground state [19,20,38,39]. Thus, the two heterostructures are designed to hold Ti chemistry approximately fixed while changing the lattice state expected to control $RuO_2$ magnetism.

Figure 1 establishes that the two heterostructures differ primarily in their epitaxial boundary condition rather than alloy composition. X-ray diffraction (XRD) $\theta$-$2\theta$ scans confirm phase-pure (110)-oriented growth in both samples, with the $SnO_2$ diffraction peak appearing only in the buffered structure. Pronounced thickness oscillations in XRD and X-ray reflectivity (XRR), together with Kiessig fringes extending to ~8° in Fig. 1(b), indicate smooth surfaces and well-defined interfaces. XRR fitting gives 2.9 nm $Ti_xRu_{1-x}O_2$/2.5 nm $TiO_2$/$TiO_2$(110) and 1.7 nm $Ti_xRu_{1-x}O_2$/1.8 nm $TiO_2$/12 nm $SnO_2$/$TiO_2$(110). Importantly, both thicknesses remain within the ultrathin regime in which coherently strained $RuO_2$/$TiO_2$ has previously exhibited an AHE [20]. More decisive for the present comparison is the alloy composition. The fitted $Ti_xRu_{1-x}O_2$ mass densities ($\rho_{alloy}$) are essentially identical, 5.732 and 5.721 g cm$^{-3}$ without and with $SnO_2$, respectively (Fig. 1d). Using $\rho_{alloy} = (1 - x)\rho_{RuO2} + x\rho_{TiO2}$, gives $x = 0.45$ and 0.46. Thus, within the sensitivity of XRR, the two films have nearly identical Ti/Ru ratios, strongly constraining composition as an explanation for any qualitative difference in Hall response.

We next verify that the two structures impose contrasting lattice states. In-situ reflection high-energy electron diffraction (RHEED) acquired immediately after growth (Fig. S3 [55]) remains sharp and streaky for the film grown directly on $TiO_2$ [Fig. 2(a)], consistent with a smooth coherently constrained surface. The $SnO_2$-buffered film instead shows broadened, diffuse streaks [Fig. 2(b)], consistent with lattice relaxation. Reciprocal-space maps (RSMs) around the asymmetric $TiO_2$ (310) and (332) reflections [Figs. 2(c)-2(f)] make this contrast quantitative. Around (332), the unbuffered $Ti_xRu_{1-x}O_2$ peak is aligned in-plane with $TiO_2$ along [001], whereas the buffered structure shifts toward the bulk $SnO_2$ position. The (310) map shows the same trend along $[1\bar{1}0]$: the unbuffered film shares the $TiO_2$ in-plane reciprocal vectors, while the buffered heterostructure approaches the bulk $SnO_2$ lattice parameter. Because the ultrathin $Ti_xRu_{1-x}O_2$ contribution cannot be separately resolved in the buffered RSM, the exact alloy lattice parameter is not directly extracted.

The crucial control is that this change in lattice state does not appreciably change the Ti valence population. Ti $2p_{3/2}$ X-ray photoelectron spectra [Figs. 2(g) and 2(h)] are well described by components assigned to $Ti^{3+}$ (~456.5 eV) and $Ti^{4+}$ (~458.0 eV). Quantitative fitting gives $Ti^{3+}$ fractions of 45.50% in the coherently strained film and 43.45% in the relaxed film. Thus, the two structures contain nearly identical, substantial $Ti^{3+}$ populations despite their markedly different strain state. Because the ~1.5 nm inelastic mean free path of Al $K_\alpha$ photoelectrons is comparable to the film thickness, the measured signal is dominated by the $Ti_xRu_{1-x}O_2$ layer. Combined with the essentially identical alloy densities and $x = 0.45$-$0.46$ compositions in Fig. 1, these measurements establish the controlled comparison needed to distinguish $Ti^{3+}$-related effects from those of strain.

The Hall response then provides a direct discrimination between the two mechanisms (Fig. 3). A pronounced nonlinearity in the field-dependent Hall resistivity, $\rho_H(H)$, develops only in the coherently strained film below 5 K, with a lower onset temperature than that reported for fully strained $RuO_2$ [20], whereas the relaxed film remains nearly linear over the full temperature range. At 1.8 K, decomposition of $\rho_H(H)$ into ordinary and anomalous contributions (see Supplemental Material [55]) yields a sizable AHE only for the strained film [Fig. 3(c)]; in the relaxed heterostructure the anomalous component is strongly suppressed [Figs. 3(d) and S4 [55]]. The key point is the controlled contrast: changing the strain state switches the anomalous Hall response while the $Ti^{3+}$ concentration changes only from 45.5% to 43.5% and the fitted alloy density remains essentially unchanged. $Ti^{3+}$ is therefore not sufficient to generate the observed anomalous Hall state. Instead, the strain selectivity is consistent with calculations in which the $TiO_2$-imposed distortion stabilizes a magnetic $RuO_2$ electronic structure [19,20,38,39]. The AHE conductivity of strained $Ti_xRu_{1-x}O_2$ is ~0.35 S $cm^{-1}$, approximately one order of magnitude smaller than in fully strained $RuO_2/TiO_2$ of comparable thickness [20]. Rather than enhancing the AHE, strong Ti incorporation therefore appears to weaken the $RuO_2$-derived response. This behavior is also distinct from that reported in doped $TiO_2$ heterostructures, where the reported AHE conductivity is much smaller (~$10^{-2}$-$10^{-8}$ S $cm^{-1}$), decreases on cooling, becomes undetectable below ~100 K, and is accompanied by pronounced ferromagnetic hysteresis [61].

If strain controls the AHE through the itinerant $RuO_2$ electronic structure, the AHE-active state should also carry a distinct metallic response. Spectroscopic ellipsometry provides such evidence [62,63]. Figures 4(a) and 4(b) show the real ($\varepsilon_1$) and imaginary ($\varepsilon_2$) parts of the dielectric function measured along $[1\bar{1}0]$ and [001]. All films exhibit an epsilon-near-zero (ENZ) energy, ($\omega_{ENZ}$), where $\varepsilon_1 = 0$, characteristic of a metallic carrier response [64]. In the coherently strained film, $\omega_{\mathrm{ENZ}}$ shifts to higher energy along both crystallographic directions and the directional splitting is reduced. The increase in $\omega_{\mathrm{ENZ}}$ corresponds to a larger effective plasma frequency, consistent with increased carrier density and/or reduced effective mass, while the reduced splitting signals a less anisotropic itinerant response. The accompanying evolution of $\varepsilon_2$ is constrained by the Kramers-Kronig relations. The loss function, $\mathrm{Im}[-1/\varepsilon(\omega)] = \varepsilon_2/(\varepsilon_1^2 + \varepsilon_2^2)$, provides an independent view [Fig. 4(c)]: the plasmon peak ($\omega(\mathrm{LF}_{\mathrm{max}})$) also shifts upward and becomes less anisotropic under strain. Quantitative fitting (see Supplemental Material [55]) shows that both the carrier scattering time ($\tau$) and plasmon relaxation time ($T_1$) increase in the strained film [Fig. 4(d)], consistent with the decrease in resistance R from 526-705 Ω in the relaxed structures to 144-314 Ω in the fully strained structures. Thus, the same epitaxial constraint that activates the AHE also reorganizes the itinerant carrier dynamics, supporting an origin tied to the strain-reconstructed metallic state rather than to a fixed population of localized $Ti^{3+}$ moments [20,62,63].

In summary, we disentangle two effects that are normally inseparable at the $RuO_2/TiO_2$ interface: epitaxial strain and Ti-related electronic reconstruction. HAXPES establishes the presence of $Ti^{3+}$ at the interface, but Ti-alloyed $RuO_2$ heterostructures show that even substantially higher $Ti^{3+}$ concentrations do not by themselves account for a robust AHE. With composition and $Ti^{3+}$ fractions held nearly fixed, the anomalous Hall response appears only under coherent epitaxial constraint and is strongly suppressed upon strain relaxation. The AHE-active state is simultaneously characterized by a higher-energy, less anisotropic plasmon response, longer carrier relaxation times, and lower resistance. These results identify strain-driven reconstruction of the itinerant electronic structure, rather than $Ti^{3+}$ formation alone, as the dominant control of the anomalous Hall state. More broadly, they demonstrate how chemically matched but lattice-distinct heterostructures can separate competing interfacial mechanisms in correlated quantum materials.

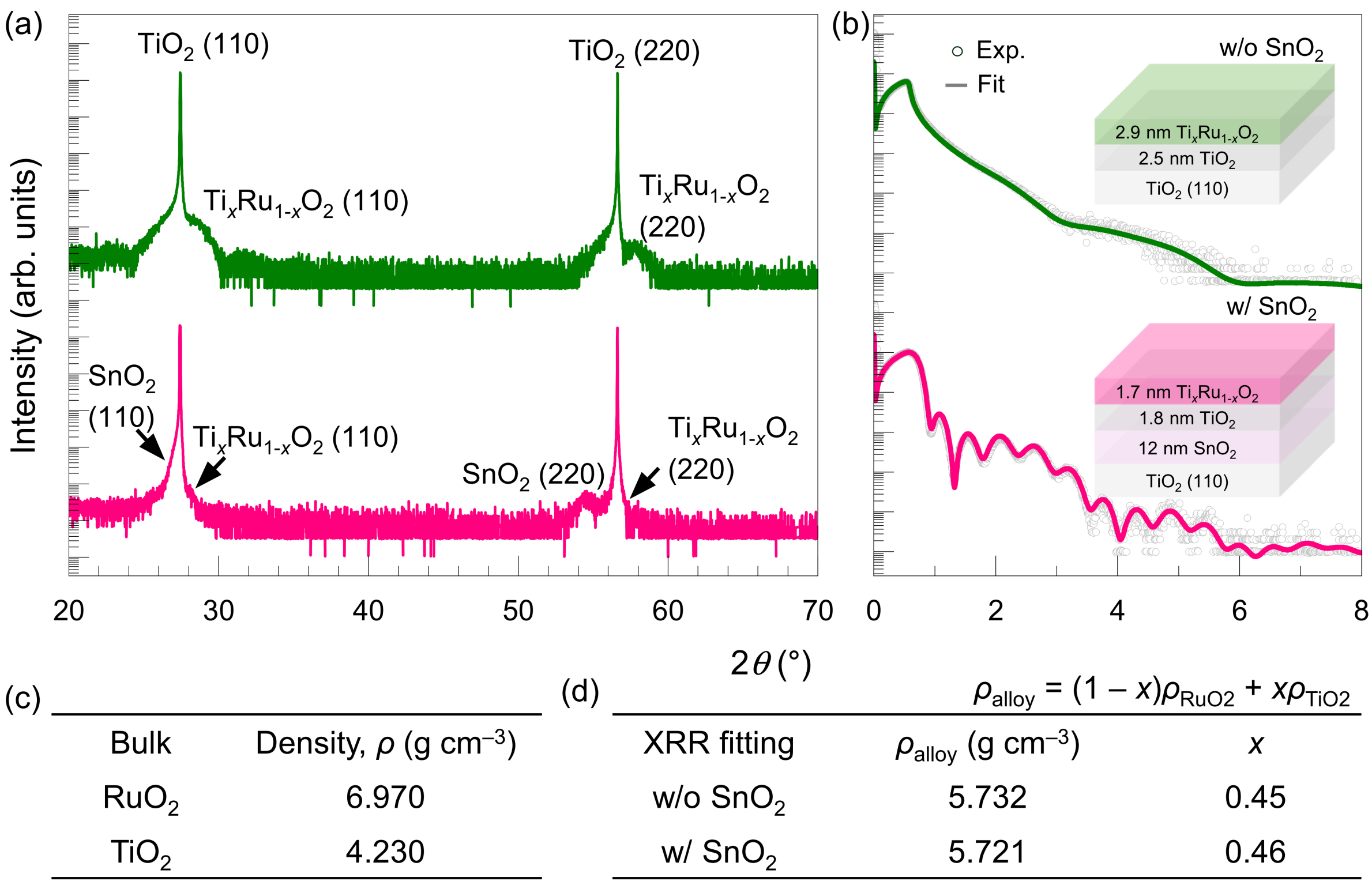

(c)

| Bulk | Density, $\rho$ (g cm$^{-3}$) |
|---|---|
| $RuO_2$ | 6.970 |
| $TiO_2$ | 4.230 |

(d) $\rho_{alloy} = (1 - x)\rho_{RuO2} + x\rho_{TiO2}$

| XRR fitting | $\rho_{alloy}$ (g cm$^{-3}$) | $x$ |
|---|---|---|
| w/o $SnO_2$ | 5.732 | 0.45 |
| w/ $SnO_2$ | 5.721 | 0.46 |

**Fig. 1. Structural and compositional characterization of epitaxial Ti-alloyed $RuO_2$ heterostructures.** (a) XRD $\theta$–$2\theta$ scans of $Ti_xRu_{1-x}O_2$ films grown on $TiO_2$ (110) without (top) and with (bottom) a $SnO_2$ buffer, showing phase-pure (110) orientation. (b) XRR data (symbols) and model fits (solid lines). (c) Bulk mass densities ($\rho$) of $RuO_2$ and $TiO_2$ used for composition estimation. (d) $\rho_{alloy}$ extracted from XRR fitting and corresponding Ti compositions ($x$), determined using a linear mixture relation (see text).

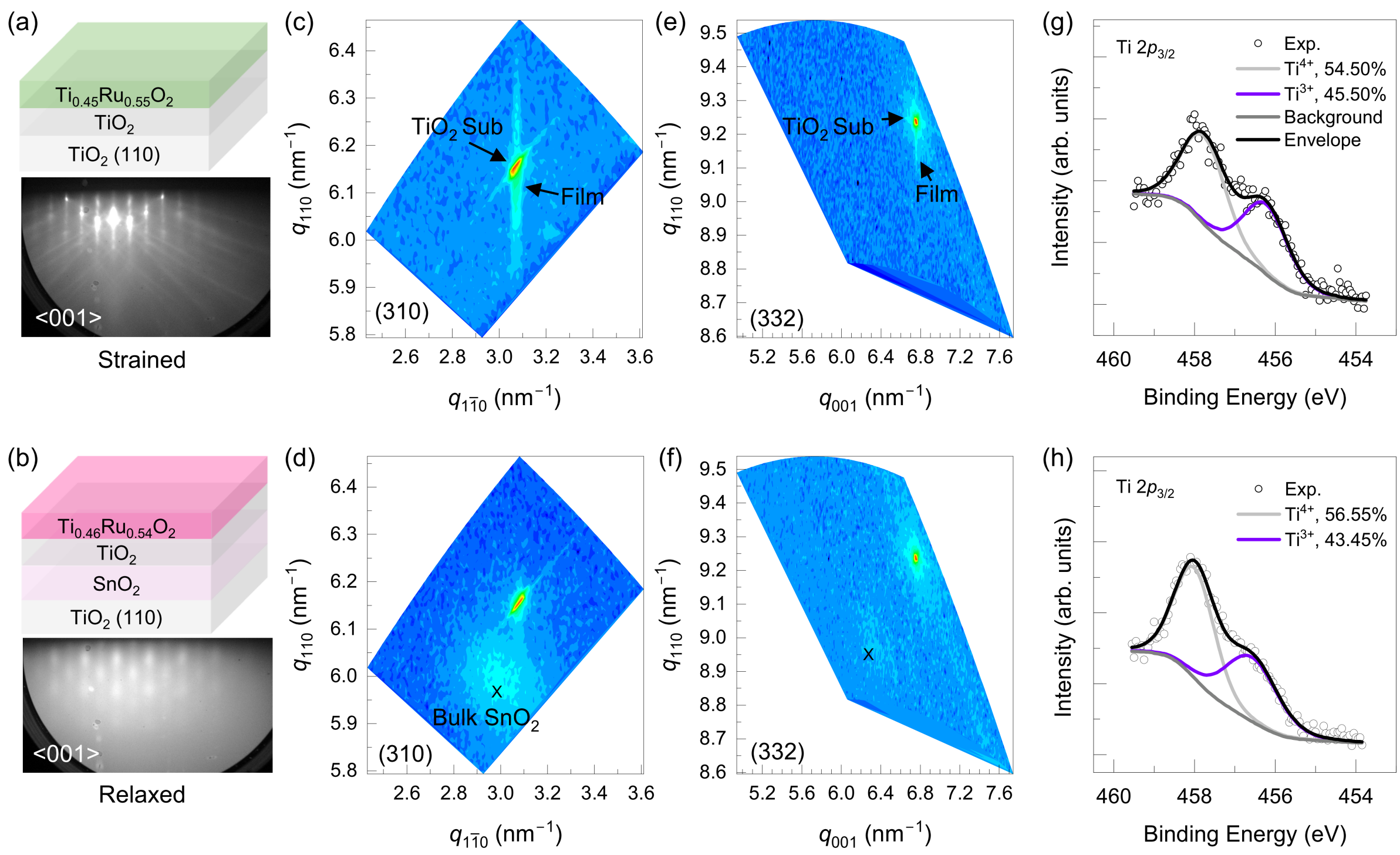

**Fig. 2. Strain state and Ti valence analysis of Ti-alloyed $RuO_2$ heterostructures.** (a,b) Schematic structures and corresponding RHEED patterns along the <001> azimuth for fully strained (without $SnO_2$ buffer) and strain-relaxed (with $SnO_2$ buffer) films. (c,d) RSMs around the (310) reflection. The film peak aligns with the $TiO_2$ substrate along in-plane $[1\bar{1}0]$ without the buffer, indicating coherent strain, while the buffered sample approaches the bulk $SnO_2$ lattice. (e,f) RSMs around the (332) reflection confirming strain along the [001] direction. (g,h) Ti $2p_{3/2}$ XPS spectra with $Ti^{4+}$ and $Ti^{3+}$ components. Both samples exhibit comparable $Ti^{3+}$ fractions.

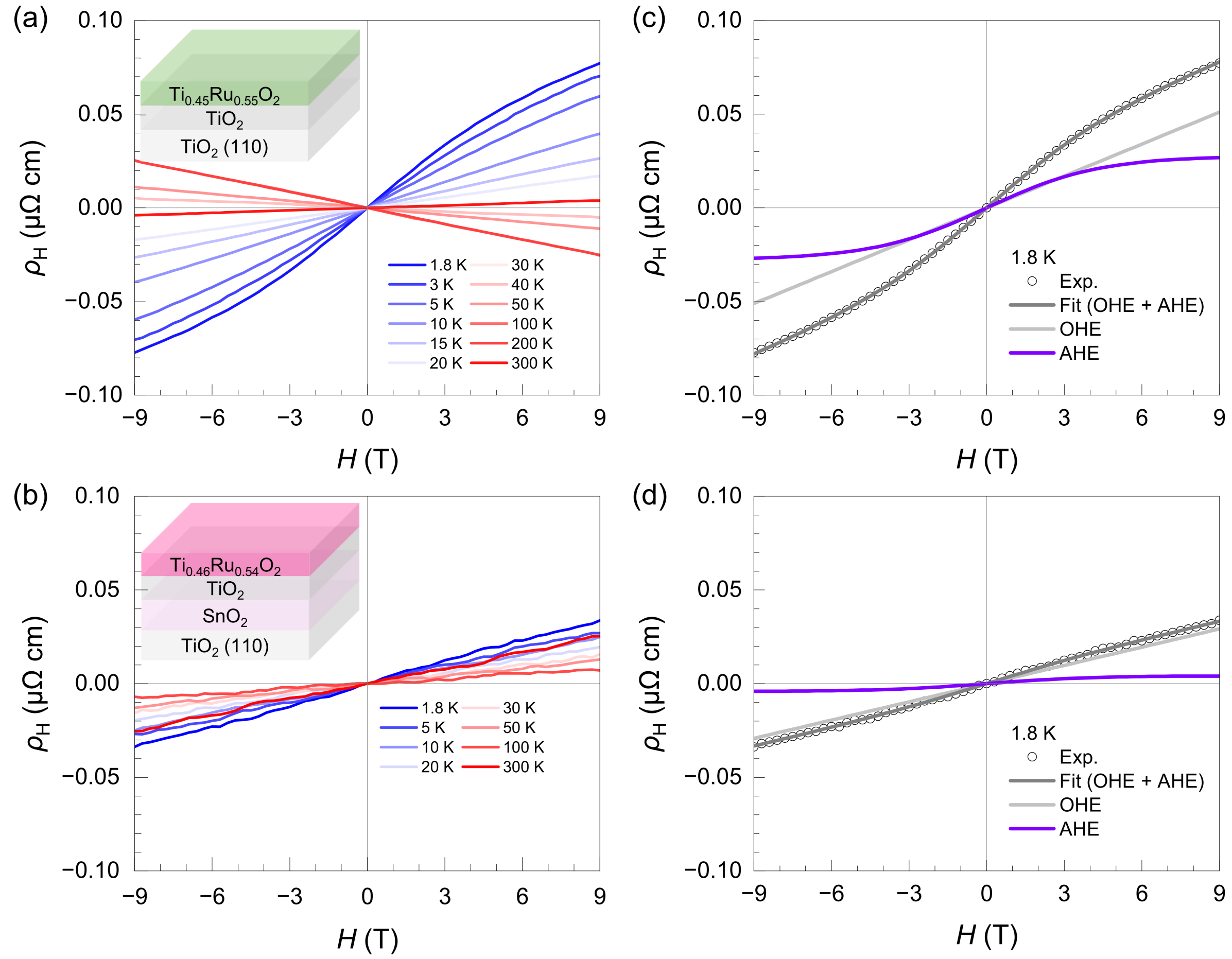

**Fig. 3. Strain-dependent anomalous Hall effect in Ti-alloyed $RuO_2$ heterostructures.** (a,b) $\rho_H(H)$ measured at various temperatures for coherently strained (a) and strain-relaxed (b) films. A pronounced nonlinearity develops only in the strained state. (c,d) Decomposition of $\rho_H$ at 1.8 K into ordinary (OHE) and anomalous (AHE) Hall contributions. Symbols denote data; lines show the total fits and extracted OHE (gray) and AHE (purple) components.

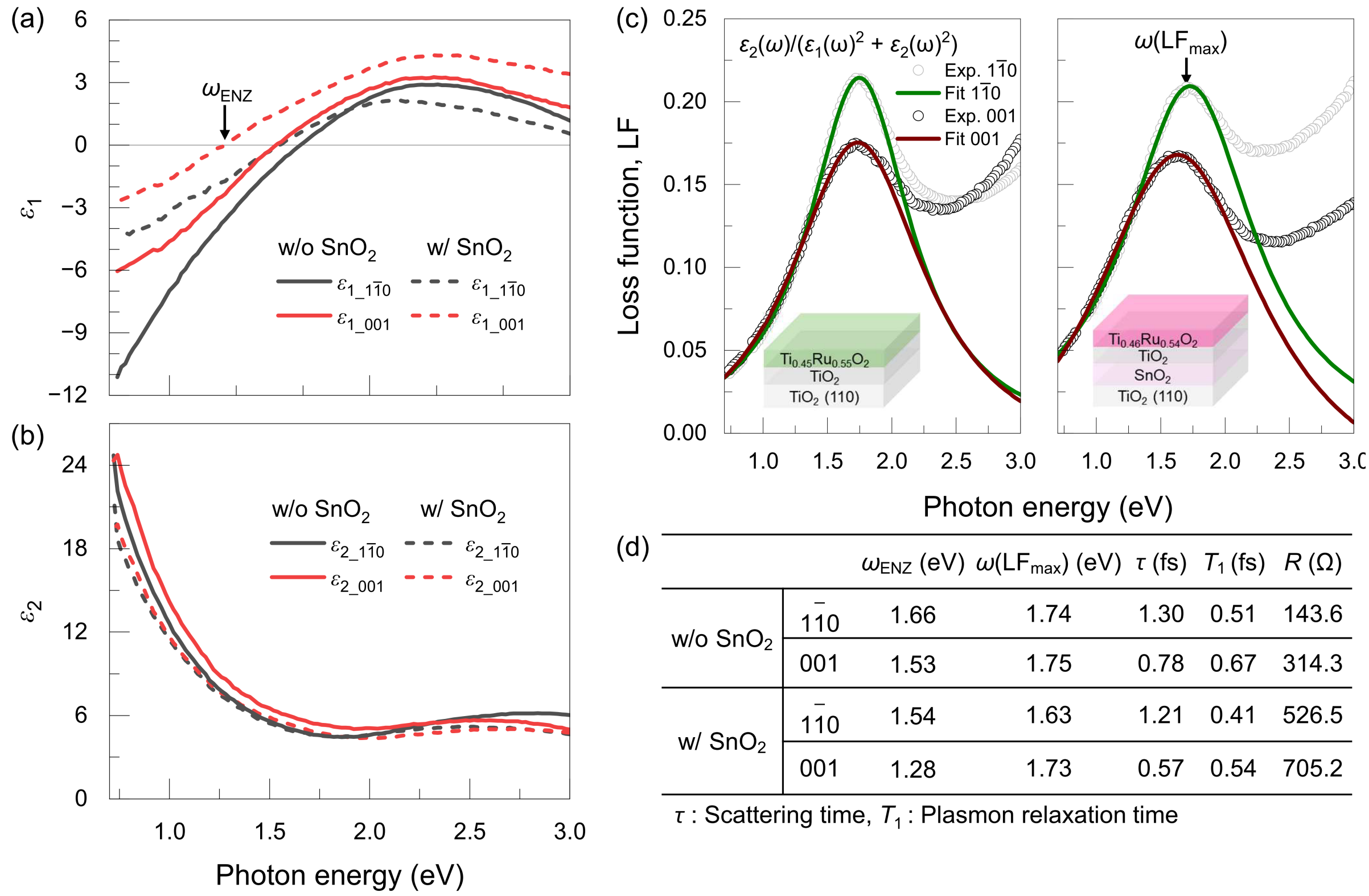

| | | $\omega_{ENZ}$ (eV) | $\omega(LF_{max})$ (eV) | $\tau$ (fs) | $T_1$ (fs) | $R$ (Ω) |
|---|---|---|---|---|---|---|
| w/o $SnO_2$ | $1\bar{1}0$ | 1.66 | 1.74 | 1.30 | 0.51 | 143.6 |
| | 001 | 1.53 | 1.75 | 0.78 | 0.67 | 314.3 |
| w/ $SnO_2$ | $1\bar{1}0$ | 1.54 | 1.63 | 1.21 | 0.41 | 526.5 |
| | 001 | 1.28 | 1.73 | 0.57 | 0.54 | 705.2 |

$\tau$ : Scattering time, $T_1$ : Plasmon relaxation time

**Fig. 4. Strain-dependent optical response and carrier dynamics.** (a,b) Real ($\varepsilon_1$) and imaginary ($\varepsilon_2$) dielectric functions from spectroscopic ellipsometry measured along $[1\bar{1}0]$ and [001]. (c) Loss-function spectra with Lorentzian fits, showing the strain-dependent plasmon response. (d) Summary of $\omega_{ENZ}$, $\omega(LF_{max})$, $\tau$, $T_1$, and $R$ at 300 K.

## Acknowledgements

Film synthesis and structural characterizations (S.G.J. and B.J.) were supported by the U.S. Department of Energy through grant Nos. DE-SC0020211, and (partly) DE-SC0024710. Transport, and ellipsometry (at UMN) were supported by the Air Force Office of Scientific Research (AFOSR) through Grant No. FA9550-21-1-0025 and FA9550-24-1-0169. Film growth was performed using instrumentation funded by AFOSR DURIP awards FA9550-18-1-0294 and FA9550-23-1-0085. S.G.J and B.J also benefitted from the support from the Air Force Office of Scientific Research Multi University Research Initiative (AFOSR MURI, Award No. FA9550-25-1-0262). S.S., U.M.J., A.X.G, S.L., and T.L. acknowledge support from AFOSR MURI, Award No. FA9550-25-1-0262. Parts of this work were carried out at the Characterization Facility, University of Minnesota, which receives partial support from the NSF through the MRSEC program under Award No. DMR-2011401. S.G.J also acknowledges support from Hankuk University of Foreign Studies Research Fund. S.A.C. was supported by the U.S. Department of Energy, Office of Science, Basic Energy Sciences, Materials Sciences and Engineering Division, Synthesis and Processing Science program (FWP 10122 at PNNL). The angle-dependent HAXPES was carried out at Diamond Light Source on beamline I09.

## Notes

The authors declare no conflict of interest.

## Data availability

The data are available from the authors upon reasonable request.